%% file: text.tex
\documentclass[showpacs,aps,superscriptaddress,pra,twocolumn,floatfix,nofootinbib]{revtex4-1}

\usepackage{graphicx}
\usepackage{color}
\usepackage[colorlinks]{hyperref}
\usepackage{subfigure}
\DeclareUnicodeCharacter{0308}{\"}

\hypersetup{
    breaklinks=false, 
    linkcolor=blue,  
    urlcolor=blue,
    citecolor=blue,
}

\usepackage{amssymb}
\usepackage{amsmath}

\begin{document}

\title{Trajectory effects on charge exchange and energy loss in collisions of H$^+$, He$^{2+}$, Li$^{3+}$, and Be$^{4+}$ ions with atomic hydrogen}

\author{Miranda Nichols}
\email{miranda.nichols@physics.gu.se}
\affiliation{Department of Physics, University of Gothenburg, SE-412 96 Gothenburg, Sweden}
\affiliation{Instituto de Ciencias Físicas, Universidad Nacional Autónoma de México, Ap. Postal 43-8, Cuernavaca, Morelos, 62251, México}

\author{D. Hanstorp}
\affiliation{Department of Physics, University of Gothenburg, SE-412 96 Gothenburg, Sweden}

\author{R. Cabrera-Trujillo}
\email{trujillo@icf.unam.mx}
\affiliation{Instituto de Ciencias Físicas, Universidad Nacional Autónoma de México, Ap. Postal 43-8, Cuernavaca, Morelos, 62251, México}
\affiliation{Department of Physics, University of Gothenburg, SE-412 96 Gothenburg, Sweden}

\begin{abstract}
The charge exchange and energy loss processes provide insights into fundamental processes across physical, chemical, and engineered systems. 
While this field has been thoroughly investigated, a clear study on trajectory effects is lacking, particularly in the context of inelastic processes at low energies.   
In this work, we address this gap by solving the time-dependent Schr\"odinger equation for the electron in a numerical lattice with a coupled electron-nuclear dynamics approach as well as a straight-line trajectory approximation for the nuclei to asses trajectory effects on charge exchange and energy loss.
The collision dynamics are studied using bare ion projectiles with charge $Z=$1-4 incident on atomic hydrogen in an energy range of 0.1 to 900 keV/u.
We find that the charge exchange process is trajectory-independent within the energy range considered, showing excellent agreement with available experimental and theoretical data. 
However, projectile energy loss exhibits strong trajectory dependence, with the straight-line approximation overestimating energy loss at low collision energies due to the forced linear path.
Our results for electronic stopping cross sections with electron-nuclear coupled trajectories are consistent with the available experimental data at high collision energies.
At low energies, nuclear energy loss becomes prominent, driven by polarization effects induced by the ion charge on the hydrogen target. 
Overall, our work highlights the importance of nuclear trajectory considerations in collision dynamics and offers a foundation for further investigations of more complex systems.
\end{abstract}

\maketitle

\section{Introduction}

Ion-atom collisions have been researched for over a century due to their role in understanding fundamental atomic structure \cite{Rutherford, Bohr}. 
Bare ions, which are completely stripped of electrons, are the simplest systems to conduct these studies with because they lack the complexity of multi-electron systems. 
In a collision, many inelastic processes can occur such as the transfer of an electron from the target to the projectile. 
This is called charge exchange or electron capture, and it plays a central role in the neutralization process \cite{Heisenberg}. 

The charge exchange process has been studied in many fields such as plasma and fusion research, astrophysics, and medicine. 
The necessity for diagnostics in plasma confinement devices such as tokamaks propelled the early interest for electron capture reactions, especially in fully stripped low-$Z$ ions \cite{Isler1994AnDiagnostic, Fonck1984DeterminationSpectroscopy}. 
Solar wind, rich in protons and alpha particles, was found to drive charge exchange processes responsible for characteristic X-ray emissions from comets \cite{Cravens1997CometIons}, planetary atmospheres \cite{Carter2012ExosphericXMM-Newton}, and even in regions of the interstellar medium \cite{Dennerl2010ChargeReactions}. 
Additionally, charge exchange plays an important role in ion beam radiation therapy for cancer treatment \cite{Champion2015WaterRadiotherapy}, particularly in the context of proton impact on hydrogen for water dissociation, where understanding radiation damage at the molecular level is crucial \cite{Luna2007Water-moleculeImpact}. 

To maximize the efficacy of these applications, it is essential to understand the physics that governs the collision dynamics. For slow and low-\textit{Z} ion-atom collisions, the atomic framework breaks down. 
In the energy region of less than 1 keV/u, a quasi-molecular representation becomes relevant,  which treats the ion and atom as a single entity due to the sharing of the electron \cite{Bransden1972TheExchange, Gaussorgues1975CommonTheory}. 
If the ion-atom pair exhibits nuclear symmetry, the lower-lying stationary states become degenerate. The H$_2^+$ molecule is the simplest example, where the electron oscillates between the resonant molecular states \cite{Lichten1963ResonantCollisions}. This is known as resonant charge transfer, which was first observed in the H$^+$+H(1$s$) system by Lockwood and Everhart \cite{Lockwood1962ResonantCollisions}. Since then, electron capture cross sections for low-\textit{Z} bare ion collisions with atomic hydrogen have been extensively studied, especially for the H$^{+}$+H(1$s$) \cite{McClure1996ElectronKeV, Hvelplund1982ElectronHydrogen, Gealy1987CrossH2, Schwab1987MeasurementHe}, He$^{2+}$+H(1$s$) \cite{Havener2005ChargeHydrogen, Shah1978ElectronCollisions, SantAnna2000ElectronicHydrogen, Hoekstra1991State-selectiveH, Basalaev2020FormationAtoms}, and Li$^{3+}$+H(1$s$) \cite{Seim-W81-14jpb3475, Shah-MB78-11jpbL233} systems. 
However, the Be$^{4+}$+H(1$s$) collision remains experimentally untouched due to the poisonous nature of beryllium.

In cases where experimental data is difficult or impossible to obtain, theoretical models are invaluable. 
These models also serve as complements to experimental results and help to contextualize findings. 
For instance, the resonant charge transfer process was investigated using the electron-nuclear dynamics (END) method for the H$^{+}$+H(1$s$) system \cite{Deumens-E94-66rmp917} to understand the importance of non-adiabatic effects for higher collision energies \cite{Killian2004ResonantEV} and for the He$^{2+}$+$^{1-3}$H systems to study isotopic effects on charge transfer \cite{Stolterfoht2007StrongTritium}. 
The END method has also been used to study charge exchange and energy loss of the proton-hydrogen collision \cite{Cabrera-Trujillo2000ChargeProjectiles}. 
In fact, the development of computational approaches to study the inelastic processes of ion-atom collisions has been significant. 
The main methods and models include solving the time-dependent Schrödinger equation (TDSE) \cite{Ludde1983DirectProbabilities, Ludde1982ElectronHydrogen}, lattice TDSE (LTDSE) \cite{Schultz1999LatticeCollisions, Minami2006, Minami2008TotalCollisions, Koakowska1999TotalKeV}, GridTDSE (GTDSE) \cite{Jorge_grid_be, Hill2023AtomicPlasmas}, basis generator method (BGM) \cite{Leung2022, Hill2023AtomicPlasmas}, classical trajectory Monte Carlo (CTMC) \cite{Olson1977Charge-transferHydrogen, Minami2008TotalCollisions,Botheron2011ClassicalCollisions,Schultz1991ProcessesPlasmas,Ziaeian2021StateEffect, Jorge2015ScalingHydrogen, Abrines1966ClassicalProtons, Elizaga1999TheoreticalCollisions, Errea2004ClassicalCollisions, Hill2023AtomicPlasmas}, atomic orbital close-coupling (AOCC) \cite{Minami2008TotalCollisions, Fritsch1982ElectronEnergies, Fritsch1984Atomic-orbital-expansionCollisions, Agueny2019ElectronStates, Toshima1994IonizationIons, Errea2006SemiclassicalFunctions, Liu2014CrossHydrogen, Bransden1983TheoreticalHe, Hill2023AtomicPlasmas}, molecular orbital close-coupling (MOCC) \cite{Botheron2011ClassicalCollisions, Harel1998CROSSRANGE, Elizaga1999TheoreticalCollisions, Errea2004ClassicalCollisions, Krstic2004CoupledApproach}, and other close-coupling treatments such as the Sturmian-pseudostate \cite{Winter2007ElectronBasis}, hidden crossing (HCCC) \cite{Havener2005ChargeHydrogen, Janev1996State-SelectiveBe4+, Krstic2004CoupledApproach}, hyperspherical (HSCC) \cite{Le2004ComparisonEnergies}, common-reaction-coordinates (CRC) \cite{Le2004ComparisonEnergies}, wave-packet convergent (WP-CCC) \cite{Antonio2021, Hill2023AtomicPlasmas}, and two-center quantum mechanical convergent (QM-CCC) \cite{Abdurakhmanov2016SolutionMethod} methods. 
There are also perturbative methods which have been shown to be more successful for high energy collisions than most of the classical and semi-classical methods mentioned above. These include various distorted wave approximations such as boundary-corrected continuum-intermediate-state method (BCIS) \cite{Milojevic2020Three-bodyMILOJE}, unitarized-distorted-wave-approximation (UDWA) \cite{Ryufuku1982IonizationHydrogen}, and continuum-distorted wave (CDW) \cite{SantAnna2000ElectronicHydrogen}) models. Other perturbative approaches include adiabatic \cite{Lichten1963ResonantCollisions}, first Born \cite{Errea2006SemiclassicalFunctions}, and eikonal impulse approximations \cite{Jorge2015ScalingHydrogen}, as well as the Bohr-Lindhard model \cite{Ding2009ClassicalAtom}. 

While there are extensive methods for studying ion-atom collisions, the primary theoretical works for calculating electron capture cross sections of bare ion projectiles include the (L)TDSE, AOCC, and CTMC methods. 
The first application of the TDSE method used a two-center-basis representation for fully stripped ions with charge $Z$=1-5, done by Lüdde and Dreizler \cite{Ludde1983DirectProbabilities, Ludde1982ElectronHydrogen}. 
However, the method required a large number of basis states, which made initial computations expensive. 
Later, the LTDSE solution was developed to bridge gaps between theoretical results and offered a level of precision that compared to that of the leading models including AOCC and MOCC, despite no use of basis sets \cite{Schultz1999LatticeCollisions}. 
Among the AOCC approaches, methods used by Fritsch and Lin (AO expansion) (AO+) \cite{Fritsch1982ElectronEnergies, Fritsch1984Atomic-orbital-expansionCollisions}, Lin \textit{et al.} (two-center (2C)-AOCC) \cite{Liu2014CrossHydrogen}, and Toshima (AOCC) \cite{Toshima1994IonizationIons} have been successfully applied to He$^{2+}$, Li$^{3+}$, and Be$^{4+}$ projectiles for a wide range of energies. 
A comparative study by Minami \textit{et al.} \cite{Minami2008TotalCollisions} evaluated the LTDSE, AOCC, and CTMC methods in an energy range of $\sim 1$ to 1000 keV/u by comparing with experimental values for total and state-selective cross sections for electron capture in the He$^{2+}$+H(1$s$) collision. They found that while both LTDSE and AOCC agree well with experiment in the entire energy range, LTDSE is ideal for low energies where the atomic orbital framework breaks down. They also note that CTMC gains reliability for high energies and capture into higher $n$ levels, which can be classically described.  
Previously considered less accurate for low energy collisions, the CTMC method has seen recent developments. 
Work by Ziaeian and Tőkési on the Be$^{4+}$+H(1$s$) system demonstrated that accounting for quantum behavior significantly improves the accuracy of CTMC, rivaling semi-classical models \cite{Ziaeian2021StateEffect}. 
 
It is clear that theoretical studies within the field of atomic collision dynamics are plentiful, yet most rely on straight-line trajectories and overlook potential trajectory effects. In this work, we address this gap by investigating energy transfer in collisions by comparing the straight-line approximation with dynamic trajectories that account for electron-nuclear coupling. Our approach utilizes LTDSE solutions to evaluate the efficacy of the method with fully stripped ions. 
Motivated by the recent straight-line trajectory calculations for energy deposition within a time-dependent density functional theory (TDDFT) framework \cite{tddft1,tddft2,tddft3,tddft4}, we believe that settling the question of trajectory effects is necessary. 
 
We focus on charge transfer and energy deposition processes in collisions of H$^{+}$, He$^{2+}$, Li$^{3+}$, and Be$^{4+}$ ions with the ground state of atomic hydrogen over an energy range of 0.1 to 900 keV/u. 
By comparing trajectory calculation methods, we evaluate the role of nuclear motion and identify the limitations of the straight-line approximation in these systems. 
Our method enables precise convergence at each time step within the lattice structure, so that our results can serve as a benchmark not only for the current studied systems but also for more complex multi-electron systems.

\section{Theory: Lattice approach}
\label{sec:lattice}
In this study, we employ a time-dependent electron-nuclear dynamics (END) approach \cite{Deumens-E94-66rmp917} for coupling electrons and nuclei dynamics. 
The time-dependent Schr\"odinger equation governing the behavior for an electron in the presence of $N$ nuclei is expressed as
\begin{equation}
\label{eq:TDSE}
  i \frac{\partial}{\partial t} \Psi({\bf r},t)= \Big[-\frac{1}{2}\nabla^2 + \sum_{i=1}^N \frac{Z_i}{|{\mathbf r}-{\mathbf R}_i(t)|}\Big] 
\Psi({\bf  r},t), \,
\end{equation}
where $\Psi({\bf r},t)$ is the time-dependent wave function, $Z_i$ is the charge of the $i$-th nuclei, and ${\bf R}_i$ denotes the trajectory of the heavy $i$-th nuclei coupled to the electron. 
The dynamics of the coupling are described by the equations \cite{Cabrera-Trujillo2023IonizationHydrogen}:
\begin{eqnarray}
\label{eq:R}
  \dot{{\mathbf R}}_i&=&\frac{1}{M_i}{\mathbf P}_i(t) \nonumber \\
  \dot{{\mathbf P}}_i&=&Z_i\langle \Psi|\frac{({\mathbf r}-{\mathbf R}_i)}{|{\mathbf r}-{\mathbf R}_i|^3}|\Psi\rangle - \\
   &  & \sum_{k\neq i}^N Z_kZ_i \frac{({\mathbf R}_k-{\mathbf R}_i)}{|{\mathbf R}_k-{\mathbf R}_i|^3}, \, \nonumber
\label{trajectory}
\end{eqnarray}
where $i$ ranges from 1 to $N$ number of heavy nuclei. 
In our specific case, we set $N=2$. 
For scenarios involving straight-line trajectories, we simplify our approach by fixing the position of the target nucleus at $\mathbf{R}_1=0$ and defining the position of the projectile as
\begin{equation}
\label{eq:straight}
\mathbf{R}_2=\mathbf{b}+\mathbf{v}t.
\end{equation}
Here, $\mathbf{b}=b\hat{x}$ represents the impact parameter and $\mathbf{v}=v_0\hat{z}$ is the collision velocity. 
In this case, the projectile momentum $\mathbf{P}_2=M_2 \mathbf{v}$ remains constant throughout the collision process and the projectile velocity is proportional to the collision energy.

\begin{figure}[t!]
\includegraphics[width=245pt]{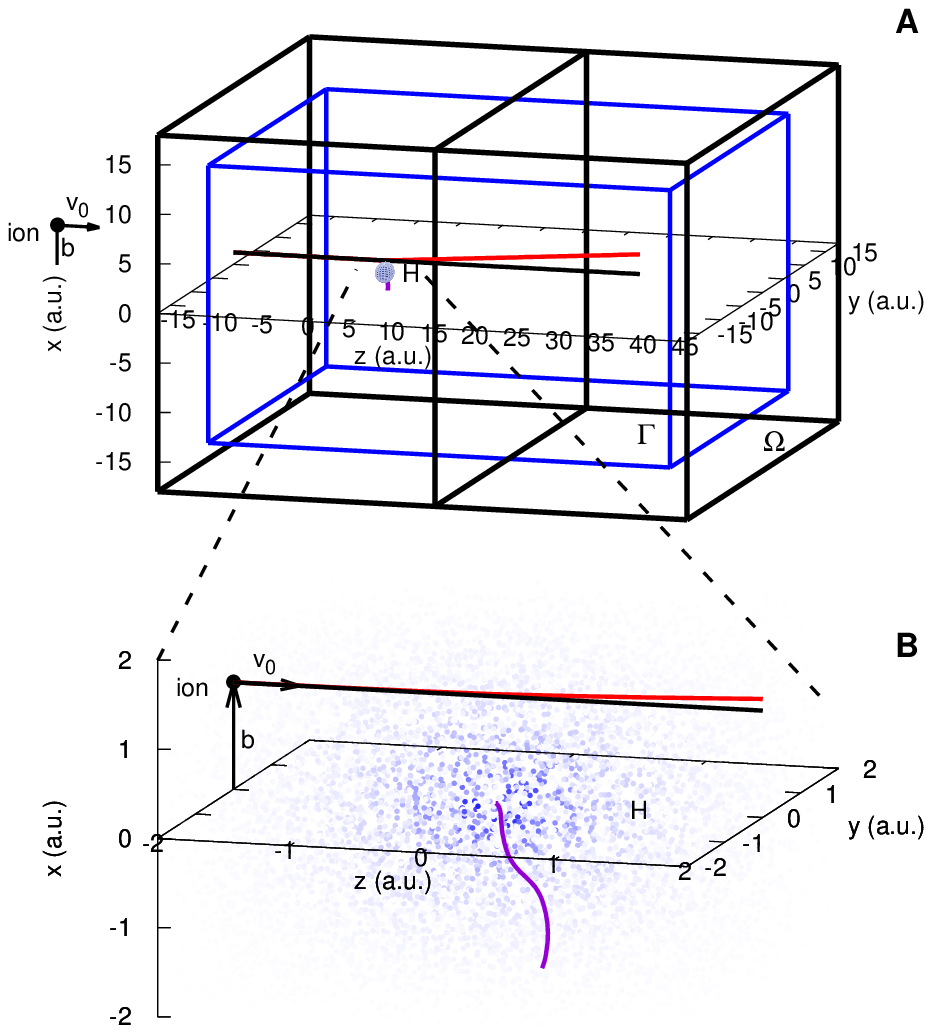}
\caption{\label{fig:1} A$^q$+H collision rendering. 
In {\bf A}, the sphere represents the $1s$ hydrogen electron density, $b$ is the collision impact parameter of an ion A$^q$ with $q$ the initial charge, located initially at a distance $z_0$, and $v_0$ is the initial collision velocity. 
The space between the inner (blue) and outer (black) boxes represents the ionization absorbing region, $\Omega$. 
The box for $z>15$ a.u. determines the electron capture by the projectile and is denoted by $\Gamma$.
The black and red lines correspond to the straight-line and the electron-nuclei coupled trajectories, respectively, for $b=1.2$ a.u. impact parameter.
In {\bf B}, an inset of the collision region around the target electronic density (dotted blue cloud) is shown.
The purple line is the target recoiling trajectory in the electron-nuclei coupled scheme.}
\end{figure}

To numerically solve Eq.~(\ref{eq:TDSE}), we utilize a lattice approach. 
A three-point finite-differences method within the Crank-Nicholson approach \cite{Crank1947AType} is applied to a uniform grid, as shown in  \cite{Fatima-A06-73pra043414}.
We address ionization by implementing a masking function that absorbs ionized electrons at each time step on the grid boundary \cite{Giovannini_2015}. 
In Fig. \ref{fig:1}, the masking function acts over the box $\Omega$ at the edge of the numerical grid in the space between the blue and black boxes. 
It is expressed as $M({\mathbf r})=M(x)M(y)M(z)$, where $M(z)$ is given by
\begin{equation}
\label{eq:masking}
M(z) = \left\{\begin{array}{lc}
\cos^{1/8}\left(\frac{\pi |z-z_{b}+L|}{2L}\right), & |z_{b}-z|<L \\
1, & \mathrm{otherwise.}\\
\end{array}\right.
\end{equation}
Here, $z_b$ indicates the boundary along the $z$-axis, either at $z_{min}$ or $z_{max}$, and $L$ is the absorbing width of the region $\Omega$. 
The expressions are similar for the $x$ and $y$ cases.

The computational grid is defined as $[-18,18]_x\times[-18,18]_y\times[-30,45]_z$~a.u., with a grid step of 0.4~a.u., as shown in Fig. \ref{fig:1}. 
Although this results in an approximate hydrogen ground state energy of $E_H=-0.490$ a.u., it is adequate for estimating charge exchange probabilities, excitation levels, and energy loss. 
Our focus is more on capturing the shapes and behaviors of the total wave functions, which is accomplished with this grid step size.

We select a time step of 0.01 a.u. which strikes a balance between accuracy and computational efficiency. It is also long enough to achieve a clear separation between the target and projectile at the end of our simulation, which is demonstrated in Fig. \ref{fig:2}.
The collision time is defined as $t_0=2 z_0/v_0$, where $v_0$ is the initial velocity of the projectile and $z_0$ is the initial projectile distance to the target. With this, Eq. (\ref{eq:R}) is solved with the fourth-order Runge-Kutta method coupled to Eq. (\ref{eq:TDSE}). For the straight trajectories, we simply advance $\mathbf{R}_2$ according to Eq. (\ref{eq:straight}).

To prevent numerical instabilities when nuclei $\mathbf{R}_i$ pass near a grid point $\mathbf{r}$, we arrange our impact parameter grid so that both projectile and target trajectories align with points at the center of adjacent squares in the $xy$-plane \cite{Fatima-A06-73pra043414}. 
The impact parameter $b$ is varied from 0.0 to 14.0 a.u. in increments of 0.4 a.u..
Initially, we position the projectile along the $z$-axis at a distance of $z_0=-30$ a.u., as illustrated in Fig. \ref{fig:1}.

We perform calculations for collision energies in the laboratory frame of the projectile, $E_p$, ranging from 0.1 keV/u up to 900 keV/u. 
To account for polarization effects on the hydrogen electronic cloud due to initial placement at $z_0=-30$ a.u. and $x=b$, we compute the initial wave function in Eq. (\ref{eq:TDSE}) using an imaginary time technique for each configuration \cite{Fatima-A06-73pra043414}, which allows for ground state convergence. 
Finally, we set an absorbing width $L=4.0$ a.u., ensuring effective electron absorption without interfering with the collision region, $\Gamma$ \cite{Giovannini_2015}. 

\section{Results}
\label{sec:results}

\subsection{Electron capture probability}

\begin{figure}[t!]
\includegraphics[width=\columnwidth]{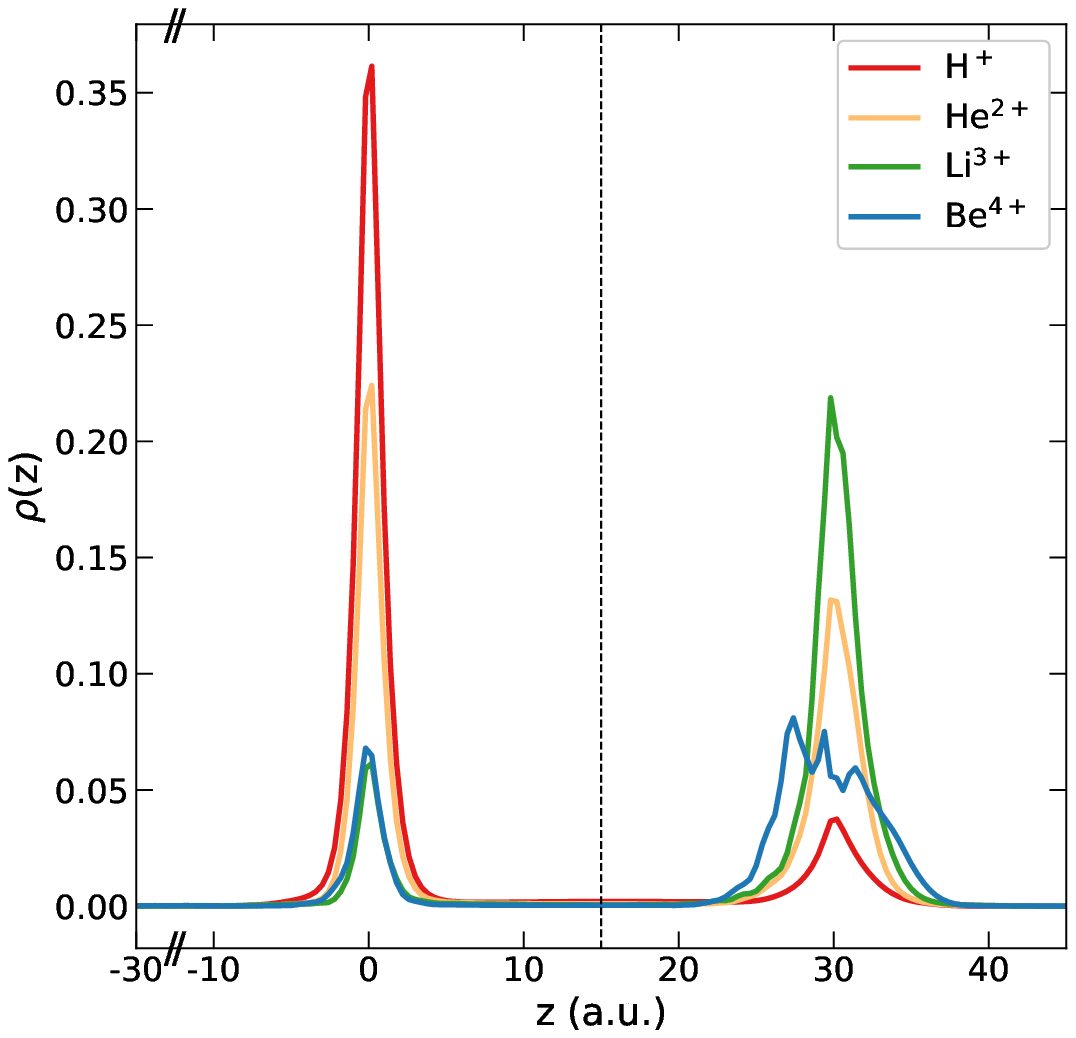}
\caption{\label{fig:2}
Electronic density as a function of $z$ when projected over $x$ and $y$ coordinates for H$^+$, He$^{2+}$, Li$^{3+}$, and Be$^{4+}$ colliding with atomic hydrogen at 5 keV/u and an impact parameter of $b=1.2$ a.u.. 
There are two regions delineated by the vertical dotted line. 
The region on the left between $-30\le z < 15$ a.u. determines the excitation region of the target. 
The region between $15\le z \le 45$ a.u. is a well-separated region after the collision where the electron capture by the projectile is determined. }
\end{figure}

In ion-atom collisions, the probability of electron capture by the projectile depends on both the impact parameter and collision energy. 
We quantify this probability using the electronic density distribution, $\rho(\mathbf{r},t) = |\Psi(\mathbf{r},t)|^2$. 
Fig. \ref{fig:2} illustrates this projected over the $xy$-plane, given as $\rho(z) = \int\int \rho(\mathbf{r},t) dx dy$, for various projectile ions (H$^+$, He$^{2+}$, Li$^{3+}$, and Be$^{4+}$) colliding with atomic hydrogen at 5 keV/u and an impact parameter of $b=1.2$ a.u.. 
The plot is divided into two regions: the left side ($-30\le z < 15$ a.u.) representing the region for excitation probability of the target, and the right side ($15\le z \le 45$ a.u.) showing the electron capture probability by the projectile ions. 

Referring to the left side of Fig. \ref{fig:2}, it can be seen that the target electronic density is notably higher for H$^+$ compared to the other ions. 
On the right side of Fig. \ref{fig:2}, we observe that the electron capture probability increases with the ionic charge, with Li$^{3+}$ exhibiting the highest peak for the given impact parameter and projectile energy. 
Interestingly, Be$^{4+}$ shows a significantly wider probability than Li$^{3+}$ despite its higher charge. 
The can be explained by considering the highest probable state populated, which is determined from the energy conservation within the system.
In this case, the initial hydrogen electronic energy should equal the final projectile electronic energy once the exchange occurs for $t\to\infty$, i.e. no projectile-target interaction. This can be summarized as 
\begin{equation}
\label{eq:conservation}
    -\frac{Z_{t}^2}{2n_i^2} = -\frac{Z_{p}^2}{2n_f^2},
\end{equation}
with $n_i$ and $n_f$ being the initial and final quantum number, in the target and projectile, respectively.
From here, one finds that $n_f=Z_p n_i/Z_t$, where the charge of the target, $Z_t$, and the initial quantum $n_i$-level are both equal to 1, therefore $n_f=Z_p$. 
This means that it is most likely for the projectile ion to capture the electron in the quantum $n$-level that matches its nuclear charge. 
In the case of Be$^{4+}$, the most probable capture will occur in the level $n$=4, which has angular momentum of $l=0$, 1, 2, and 3 corresponding to the $s$, $p$, $d$, and $f$ orbitals. 
The introduction of higher angular momentum orbitals inherently increases the complexity of the system and can explain the broader behavior of the capture probability.  

\begin{figure*}[t!]
\includegraphics[width=\textwidth]{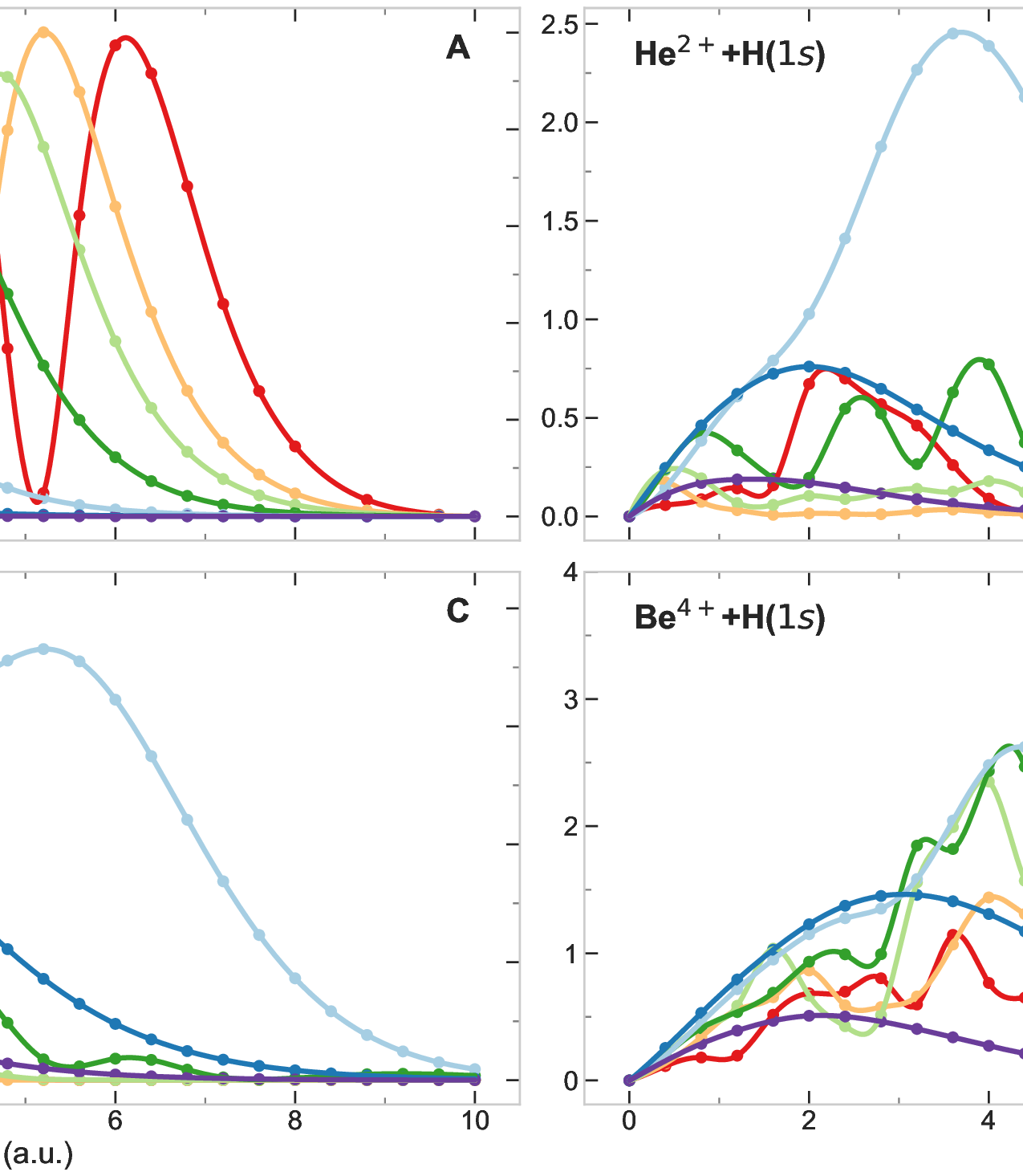}
\caption{\label{fig:3}
Electron capture probability, weighted by the impact parameter, as a function of the impact parameter for collision energies of 0.1, 0.25, 0.5, 1.5, 10, 50, and 100 keV/u. Results for ({\bf A}) H$^++$H($1s$), ({\bf B}) He$^{2+}+$H($1s$), ({\bf C}) Li$^{3+}+$H($1s$), and ({\bf D}) Be$^{4+}+$H($1s$) are shown. 
The data points correspond to our electron-nuclei coupled results of Eq. (\ref{eq:R}), and the solid lines represent cubic spline fits and serve only as a visual guide.}
\end{figure*}

\begin{figure*}[t!]
\centering
\includegraphics[width=\textwidth]{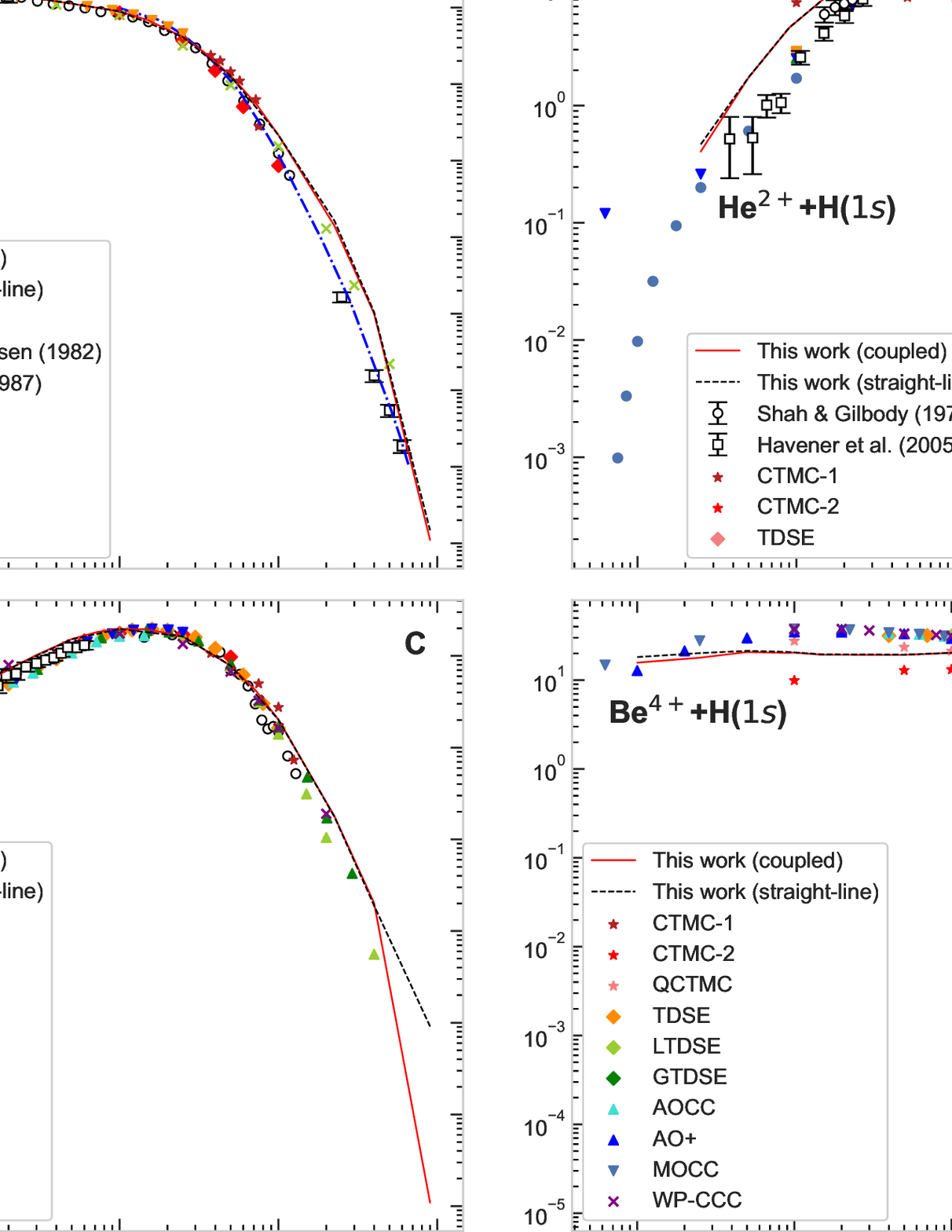}
\caption{\label{fig:4}
Electron capture cross sections for (\textbf{A}) H$^+$, (\textbf{B}) He$^{2+}$, (\textbf{C}) Li$^{3+}$, and (\textbf{D}) Be$^{4+}$ colliding with H($1s$), as a function of the ion projectile energy. 
Our results from the electron-nuclei coupled trajectories are shown with a solid red line and the straight-line trajectories are shown with a dashed black line. 
In {\bf A}, we compare with the experimental data from McClure \cite{McClure1996ElectronKeV}, Hvelplund and Andersen \cite{Hvelplund1982ElectronHydrogen}, and Gealy and Van Zyl \cite{Gealy1987CrossH2}, as well as the theoretical results from CTMC \cite{Olson1977Charge-transferHydrogen}, LTSDE \cite{Koakowska1999TotalKeV}, MOCC \cite{Harel1998CROSSRANGE}, UDWA \cite{Ryufuku1982IonizationHydrogen}, and QM-CCC \cite{Abdurakhmanov2016SolutionMethod}.
In {\bf B}, we compare with the experimental data from Shah and Gilbody \cite{Shah1978ElectronCollisions} and Havener \textit{et al.} \cite{Havener2005ChargeHydrogen}, as well as the theoretical results from CTMC-1 \cite{Minami2008TotalCollisions}, CTMC-2 \cite{Olson1977Charge-transferHydrogen}, TDSE \cite{Ludde1982ElectronHydrogen}, LTDSE \cite{Minami2008TotalCollisions}, AOCC-1 \cite{Minami2008TotalCollisions}, AOCC-2 \cite{Toshima1994IonizationIons}, AOCC-3 \cite{Bransden1983TheoreticalHe}, MOCC \cite{Harel1998CROSSRANGE}, END \cite{Stolterfoht2007StrongTritium}, and Sturmian-pseudostate \cite{Winter2007ElectronBasis}.
In {\bf C}, we compare with the experimental data from Seim \textit{et al.} \cite{Seim-W81-14jpb3475} and Shah \textit{et al.} \cite{Shah-MB78-11jpbL233}, as well as the theoretical results from CTMC \cite{Olson1977Charge-transferHydrogen}, TDSE \cite{Ludde1982ElectronHydrogen}, 2C-BGM \cite{Leung2022}, AOCC \cite{Toshima1994IonizationIons}, AO+ \cite{Fritsch1982ElectronEnergies}, 2C-AOCC \cite{Liu2014CrossHydrogen}, MOCC \cite{Harel1998CROSSRANGE}, and UDWA \cite{Ryufuku1982IonizationHydrogen}.
In {\bf D}, we compare with the theoretical results from CTMC-1 \cite{Olson1977Charge-transferHydrogen}, CTMC-2 \cite{Ziaeian2021StateEffect}, QCTMC \cite{Ziaeian2021StateEffect}, TDSE \cite{Ludde1982ElectronHydrogen}, LTDSE \cite{Minami2006}, GTDSE \cite{Jorge_grid_be}, AOCC \cite{Toshima1994IonizationIons}, AO+ \cite{Fritsch1984Atomic-orbital-expansionCollisions}, MOCC \cite{Harel1998CROSSRANGE}, and WP-CCC \cite{Antonio2021}.}
\end{figure*}

The electron capture probability is given as the integration of the electronic density in the region of the projectile well after the collision, $\Gamma$ (Fig. \ref{fig:1}), $P_{CX}=\langle \Psi(\mathbf{r},t)|\Psi(\mathbf{r},t)\rangle_{\Gamma}$. 
The analysis of weighted probabilities against impact parameter for energies ranging from 0.1 keV/u to 100 keV/u, as shown in Fig. \ref{fig:3}, provides more insights on $n$-level capture. 
As H$^+$ is most likely to capture in the $n$=1 level, the H$^+$+H(1$s$) collision in Fig. \ref{fig:3}{\bf A} demonstrates a strong resonant electron capture effect for energies below 10 keV/u. 
This tends to be stronger at lower energies due to the dominant Coulomb interaction and ease of resonant channel coupling. 
At higher energies, shorter interaction times result in lower electron capture probabilities. 
There is also no resonant charge exchange for higher energies as interaction time determines the number of oscillations. 
In other words, oscillations dampen with increasing energy due to coupling to different, non-resonant channels of the system.

The other ion collisions show varying degrees of this resonant effect. 
The He$^{2+}$+H(1$s$) collision in Fig. \ref{fig:3}{\bf B} is only slightly asymmetric and therefore exhibits resonant effects which appear the strongest at 1.5 keV/u projectile energy. 
As the ionic charge increases, so does the asymmetry of the collision, which causes the resonant channel coupling to become weaker. 
Like He$^{2+}$, the Li$^{3+}$+H(1$s$) collision in Fig. \ref{fig:3}{\bf C} also has the highest probability for 10 keV/u but diminishing resonant effects at 1.5 keV/u. 
While there still appears to be a resonant channel for the Be$^{4+}$+H(1$s$) collision in Fig. \ref{fig:3}{\bf D}, it deviates from the trends observed in the other collisions. 
Here, the maximum electron capture probability occurs at  $b$=6.4 for a collision energy of 0.25 keV/u. 
The preference for Be$^{4+}$ to capture in higher \textit{b} values can be traced back to Eq. (\ref{eq:conservation}) where $n=4$ is more extended than the other systems.

\begin{figure*}[!t]
\centering
\includegraphics[width=510pt]{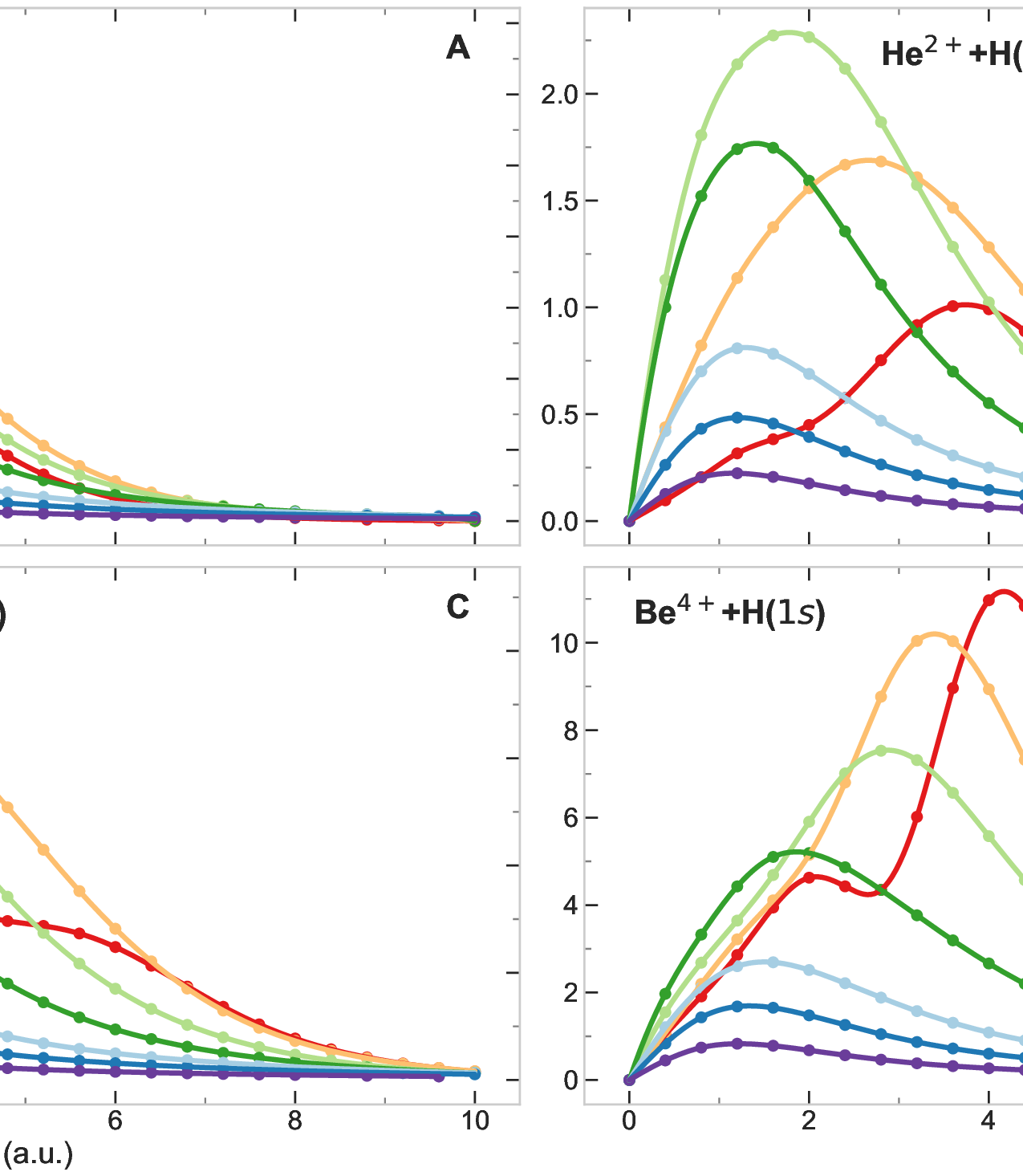}
\caption{\label{fig:5}
Projectile electronic energy loss, weighted by the impact parameter, as a function of the impact parameter for collision energies of 10, 25, 50, 100, 225, 400, and 900 keV/u..
Results for ({\bf A}) H$^++$H($1s$), ({\bf B}) He$^{2+}+$H($1s$), ({\bf C}) Li$^{3+}+$H($1s$), and ({\bf D}) Be$^{4+}+$H($1s$) are shown. 
The data points correspond to our electron-nuclei coupled results of Eq. (\ref{eq:R}), and the solid lines represent cubic spline fits and serve only as a visual guide.}
\end{figure*}

\subsection{Electron capture cross section}

The area under the electron capture probability curves of Fig. \ref{fig:3} is proportional to the electron capture cross sections, since
\begin{equation}
\label{eq:sigmai}
\sigma=2\pi\int b P_{CX}\, db\,.
\end{equation}
In Fig. \ref{fig:4}, we show the electron capture cross sections as a function of the projectile collision energy for H$^+$, He$^{2+}$, Li$^{3+}$, and Be$^{4+}$ colliding with atomic hydrogen. 
In Fig. \ref{fig:4}{\bf A}, we compare our H$^+$ results with the experimental data from McClure \cite{McClure1996ElectronKeV}, Hvelplund and Andersen \cite{Hvelplund1982ElectronHydrogen}, and Gealy and Van Zyl \cite{Gealy1987CrossH2}, as well as the theoretical results from CTMC \cite{Olson1977Charge-transferHydrogen}, LTSDE \cite{Koakowska1999TotalKeV}, MOCC \cite{Harel1998CROSSRANGE}, UDWA \cite{Ryufuku1982IonizationHydrogen}, and QM-CCC \cite{Abdurakhmanov2016SolutionMethod}. 
Due to the resonant character of the electron capture, $n_i=n_f=1$ ($s$ orbital) in Eq. (\ref{eq:conservation}), as the projectile energy is reduced, the electron capture cross section increases \cite{Lichten1963ResonantCollisions}.
We observe that our numerical results agree with the experimental data within the experimental error bars for the low and intermediate energy ranges.
Our results from both the straight-line and the electron-nuclei coupled trajectories agree well with each other. This implies that the electron capture process is not affected by the trajectory at these collision energies.
This is due to the adiabatic molecular pseudo-potential formed by the projectile and target in this energy range, which only depends on the projectile-target distance. Thus, the formation of the pseudo-potential, and in turn the electron capture cross sections, are not significantly influenced by the kinematics of the collision \cite{Lichten1963ResonantCollisions}.

In Fig. \ref{fig:4}{\bf B}, we compare our He$^{2+}$ results with the experimental data from Shah and Gilbody \cite{Shah1978ElectronCollisions} and Havener \textit{et al.} \cite{Havener2005ChargeHydrogen}, as well as the theoretical results from CTMC-1 \cite{Minami2008TotalCollisions}, CTMC-2 \cite{Olson1977Charge-transferHydrogen}, TDSE \cite{Ludde1982ElectronHydrogen}, LTDSE \cite{Minami2008TotalCollisions}, AOCC-1 \cite{Minami2008TotalCollisions}, AOCC-2 \cite{Toshima1994IonizationIons}, AOCC-3 \cite{Bransden1983TheoreticalHe}, MOCC \cite{Harel1998CROSSRANGE}, END \cite{Stolterfoht2007StrongTritium}, and Sturmian-pseudostate \cite{Winter2007ElectronBasis}. 
Here, the electron is most likely captured in $n_f=2$, which involve the $s$ and $p$ orbitals. Consequently, the electron capture cross sections for the He$^{2+}$ projectile exhibit a different trend compared to those for the H$^+$ projectile.
The trend shows a peak at 10 keV/u and decreases for lower collision energies, which slightly overestimates the experimental data but still reflects the same behavior. 

\begin{figure*}[t!]
\includegraphics[width=\textwidth]{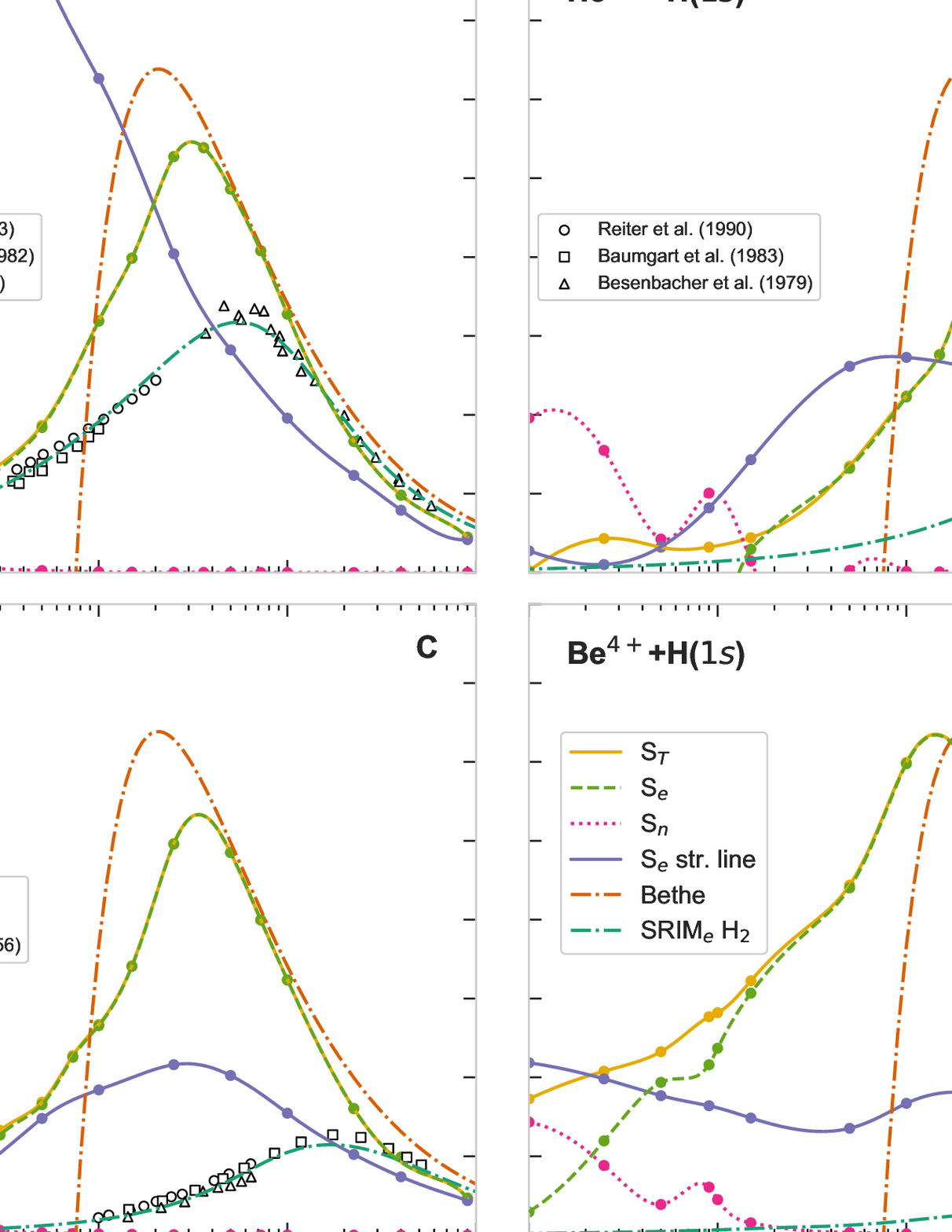}
\caption{\label{fig:6}
Stopping cross sections for atomic hydrogen targets by incident (\textbf{A}) H$^+$, (\textbf{B}) He$^{2+}$, (\textbf{C}) Li$^{3+}$, and (\textbf{D}) Be$^{4+}$ ions, as a function of the projectile energy. 
The solid goldenrod line corresponds to the total stopping cross section, $S_T$, determined from the coupled trajectories of Eq. (\ref{eq:R}).
The dashed green line is the electronic contribution, $S_e$, which is obtained by subtracting the nuclear energy loss from the total energy loss.
The dotted pink line represents the nuclear stopping cross section, $S_n$.
The solid purple line shows the electronic stopping cross section from the straight-line trajectories of Eq. (\ref{eq:St}), determined by the electronic energy gain from the target. 
The data points correspond to our results, fitted with a cubic spline as a guide. For the S$_e$ line in \textbf{B}, the data is fitted with a piecewise cubic Hermite interpolation to preserve the curve shape. 
For comparison, we plot the Bethe analytical solution (Eq. (\ref{eq:bethe})) \cite{Bethe-H30,Bethe-J86,Inokuti-M71}, with a dash-dot dark-orange line. The dash-dot dark-green line represents the electronic stopping cross sections for H$_2$ targets by incident (\textbf{A}) H, (\textbf{B}) He, (\textbf{C}) Li, and (\textbf{D}) Be ions from SRIM \cite{SRIM}. Select experimental data, compiled in Helmut Paul's database \cite{Helmut-Paul,Montanari-CC24-551nimb165336}, are also shown for H$_2$ targets by incident (\textbf{A}) H \cite{Schiefermuller1993EnergyGases, Brgesen1982StoppingHydrogen, H&Hestopping}, (\textbf{B}) He \cite{Reiter1990PROTONSECTIONS, H.Baumgart19834HeCO2, H&Hestopping}, and (\textbf{C}) Li \cite{H.H.Andersen1978StoppingXe,stopping_teplova, Allison1956StoppingEnergy} ions. There is no available experimental data for Be ions.}
\end{figure*}

In Fig. \ref{fig:4}{\bf C}, we compare our Li$^{3+}$ results with the experimental data from Seim \textit{et al.} \cite{Seim-W81-14jpb3475} and Shah \textit{et al.} \cite{Shah-MB78-11jpbL233}, as well as theoretical results from CTMC \cite{Olson1977Charge-transferHydrogen}, TDSE \cite{Ludde1982ElectronHydrogen}, 2C-BGM \cite{Leung2022}, AOCC \cite{Toshima1994IonizationIons}, AO+ \cite{Fritsch1982ElectronEnergies}, 2C-AOCC \cite{Liu2014CrossHydrogen}, MOCC \cite{Harel1998CROSSRANGE}, and UDWA \cite{Ryufuku1982IonizationHydrogen}. 
In this case, the electron capture occurs mostly for states with $n_f\le 3$ involving $s$, $p$, and $d$ orbitals which result in a peak around 15 keV/u. There is a very good agreement of our numerical results to the experimental data and other theoretical approaches except for the UDWA model \cite{Ryufuku1982IonizationHydrogen} for energies of 20 keV/u and less.

There is no experimental data for the Be$^{4+}$ + H(1$s$) collision, so in Fig. \ref{fig:4}{\bf D}, we compare our results to the theoretical results from CTMC-1 \cite{Olson1977Charge-transferHydrogen}, CTMC-2 \cite{Ziaeian2021StateEffect}, QCTMC \cite{Ziaeian2021StateEffect}, TDSE \cite{Ludde1982ElectronHydrogen}, LTDSE \cite{Minami2006}, GTDSE \cite{Jorge_grid_be}, AOCC \cite{Toshima1994IonizationIons}, AO+ \cite{Fritsch1984Atomic-orbital-expansionCollisions}, MOCC \cite{Harel1998CROSSRANGE}, and WP-CCC \cite{Antonio2021}.
Interestingly, we find that the electron capture cross section remains almost constant for collision energies below 25 keV/u. This could be a consequence of the uniform distribution of angular momentum orbitals in the higher-\textit{n} capture levels. Contrary to Eq. (\ref{eq:conservation}), some \textit{n}-partial cross section studies observed that the electron is predominantly captured into the $n=3$ level for low to intermediate collision energies and the $n=2$ level for the high-energy region \cite{Jorge_grid_be, Antonio2021, Fritsch1984Atomic-orbital-expansionCollisions}. Nevertheless, the physical behavior in the energy region is consistent with the compared theoretical models.

\subsection{Projectile energy loss}
The energy required to induce the excitation and charge transfer processes comes from the kinetic energy loss of the projectile.
This is given by $\Delta E_T=K_f-K_i$, where $K_f$ and $K_i$ are the final and initial kinetic energy of the projectile. The kinetic energy is represented by $K=\mathbf{P}_2^2/2M_p$, where the final projectile momentum, $\mathbf{P}_2$, is given by Eq. (\ref{eq:R}).
The projectile energy loss is divided into two components: the energy loss within the relative system when the target at rest, and the energy loss from the displacement of the target. 
Thus, the relative energy loss is $\Delta E_e=K^r_f-K^r_i$ with $K^r=p_r^2/2\mu_p$, where $p_r$ is the relative momentum of a projectile with reduced mass $\mu=M_pM_t/(M_p+M_t)$.
The energy loss in the relative system results in excitations or ionization of the target, which is referred to as the electronic energy loss.

In Fig. \ref{fig:5}, we show the projectile electronic energy loss, weighted by the impact parameter, as a function of the initial projectile energy and impact parameter.
By comparing with the results of Fig. \ref{fig:3}, we observe a correlation between the charge exchange process and the projectile electronic energy loss. Specifically, the transfer of kinetic energy from the projectile to the target results in electronic loss within the target, which can be absorbed by the projectile. 

However, the energy loss of the projectile also contributes to displacing the hydrogen atom, causing the nucleus of the target to recoil as a result of the momentum transferred by the projectile.
As the hydrogen atom momentum is initially at rest, given by Eq. (\ref{eq:R}), the energy gained by the target as nuclear recoil is $\Delta E_n=K_t^f={\mathbf P}_1^2/2M_t$.
Therefore, the total energy loss is the sum of these two contributions, $\Delta E_T=\Delta E_e+\Delta E_n$.
At high collision energies or large impact parameters, the target recoil is negligible.
However, at low collision energies or small impact parameters, the target kinetic energy increases due to polarization effects induced by the projectile on the hydrogen target.

Once the electronic energy loss is determined, the electronic stopping cross section is given by
\begin{equation}
\label{eq:St}
S_e(E_p)=2\pi \int b \Delta E_e\,db,\,
\end{equation}
where the nuclear stopping cross section, $S_n$, is given by replacing $\Delta E_e$ with $\Delta E_n$. For the electronic energy loss in the straight-line trajectories, the constant projectile velocity results in $\Delta E_T=0$. However, it is customary to subtract the electronic energy loss of the projectile from the electronic energy gain of the target. This approach is commonly used in TDDFT methods \cite{tddft1,tddft2,tddft3, tddft4}. By calculating the final electronic energy of the target from the final wave function in the interval $-30 < z < 15$ a.u. and subtracting the initial ground state energy of the target, we isolate the straight-line contribution to the electronic energy loss. 

\setlength{\tabcolsep}{4.25pt}
\renewcommand{\arraystretch}{1.3}
\begin{table*}[!t]
\caption{\label{tab:1}
Electron capture cross section, $\sigma$ (10$^{-16}
$ cm$^2$), total stopping cross section, $S_T$ (10$^{-15}$ eV cm$^2$), and nuclear stopping cross section, $S_n$ (10$^{-15}$ eV cm$^2$), for H$^+$, He$^{2+}$, Li$^{3+}$, and Be$^{4+}$ ions colliding with H(1$s$) as a function of the projectile energy ($E$ in keV/u) for the case of electron-nuclei coupled trajectories.}
\begin{tabular}{|l| ccc | ccc | ccc | ccc |}
\hline
 & \multicolumn{3}{|c|}{H$^+$} & \multicolumn{3}{|c|}{He$^{2+}$} & \multicolumn{3}{|c|}{Li$^{3+}$} & \multicolumn{3}{|c|}{Be$^{4+}$} \\ \hline
 $E (keV/u)$  &    $\sigma$ & $S_T$ & $S_n$ & $\sigma$ & $S_T$ & $S_n$ & $\sigma$ & $S_T$ & $S_n$ & $\sigma$ & $S_T$ & $S_n$ \\  \hline
0.1   & 25.814 &  3.276 &  7.55739 &   2.332 &   0.196 & 15.6748 &   1.00315 & 32.356 &  34.8949 &  15.740981 &   55.2884 &  45.8768 \\
0.25  & 22.782 &  1.866 &  3.17662 &   0.405 &   3.461 & 12.3975 &   1.07430 & 14.775 &  20.1815 &  17.860768 &   66.5140 &  28.2684 \\
0.5   & 20.570 &  1.456 &  2.95391 &   1.739 &   2.558 &  3.3956 &   1.81666 & 12.531 &   7.5884 &  20.704486 &   74.4218 &  12.4336 \\
0.9   & 18.251 &  1.272 &  1.04834 &   4.606 &   2.582 &  8.0594 &   3.79431 &  1.832 &  18.2927 &  20.266769 &   88.6153 &  19.4791 \\
1.5   & 16.261 &  1.768 &  0.26016 &   8.226 &   3.543 &  1.1667 &   5.39327 & 16.631 &   3.1981 &  20.390969 &   90.2421 &  14.4162 \\
5.0   & 11.201 &  3.730 &  0.05163 &  15.079 &  10.790 &  0.2158 &  15.28633 & 30.334 &   0.5616 &  19.317279 &  103.2013 &   5.0087 \\
10.0  &  8.882 &  6.393 &  0.01998 &  15.395 &  17.914 &  0.1082 &  19.65515 & 48.067 &   0.2699 &  19.233950 &  141.9972 &   1.0964 \\
15.0  &  6.851 &  7.978 &  0.01115 &  14.300 &  22.127 &  0.0694 &  19.51015 & 61.494 &   0.1794 &  20.261358 &  191.7968 &   0.5270 \\
25.0  &  4.125 & 10.550 &  0.00656 &  10.912 &  34.174 &  0.0386 &  16.31505 & 89.340 &   0.1057 &  21.790332 &  191.4832 &   0.3505 \\
36.0  &  2.381 & 10.774 &  0.00456 &   7.366 &  38.883 &  0.0250 &  11.87955 & 87.918 &   0.0705 &  20.541394 &  181.0882 &   0.2076 \\
50.0  &  1.281 &  9.727 &  0.00321 &   4.514 &  40.316 &  0.0167 &   7.83414 & 87.351 &   0.0479 &  16.434467 &  152.1292 &   0.1408 \\
72.2  &  0.552 &  8.162 &  0.00226 &   2.199 &  36.219 &  0.0107 &   4.15584 & 71.937 &   0.0305 &  11.682945 &  143.3466 &   0.0971 \\
100.0 &  0.214 &  6.554 &  0.00164 &   0.986 &  27.841 &  0.0073 &   2.01806 & 58.256 &   0.0203 &   6.616649 &  119.5971 &   0.0632 \\
225.0 &  0.014 &  3.325 &  0.00073 &   0.074 &  13.237 &  0.0030 &   0.17858 & 28.884 &   0.0075 &   3.227359 &   97.8711 &   0.0425 \\
400.0 &  0.001 &  1.963 &  0.00044 &   0.008 &   8.031 &  0.0018 &   0.02020 & 17.814 &   0.0041 &   0.301962 &   49.3949 &   0.0153 \\
900.0 &  1.1e-6 & 0.897 &  0.00018 &   0.003 &   3.791 &  0.0007 &   0.00001 &  8.455 &   0.0016 &   0.034990 &   30.5431 &   0.0078 \\ \hline
\end{tabular}
\end{table*}

In Fig. \ref{fig:6}, we show the total (solid goldenrod line), electronic (dashed green line), and nuclear (dotted pink line) stopping cross sections, $S_T, S_n,$ and $S_e$, respectively, for atomic hydrogen targets by incident projectile ions. Since the target is the same and only the projectile varies, the stopping cross sections are normalized by the square of the projectile charge. 

We compare our electronic stopping cross sections with experimental data for H$_2$ targets by incident H \cite{Schiefermuller1993EnergyGases, Brgesen1982StoppingHydrogen, H&Hestopping}, He \cite{Reiter1990PROTONSECTIONS, H.Baumgart19834HeCO2, H&Hestopping}, and Li \cite{H.H.Andersen1978StoppingXe,stopping_teplova, Allison1956StoppingEnergy} ions, compiled by Helmut Paul at the IAEA \cite{Helmut-Paul,Montanari-CC24-551nimb165336}, and with SRIM \cite{SRIM} calculations also for H$_2$ targets (dash-dot dark-green line).
For further comparison, we include Bethe's analytical formula (dash-dot dark-orange line) from \cite{Bethe-H30,Bethe-J86,Inokuti-M71},
\begin{equation}
\label{eq:bethe}
S_e=\frac{4\pi e^4}{mv^2}Z_p^2 Z_t\ln{\left(\frac{2mv^2}{I_0}\right)},
\end{equation}
where $I_0$ is the target mean excitation energy.
For atomic hydrogen, $I_0=14.9$ eV is used, as reported by Bethe \cite{Bethe-J86,Inokuti-M71} and by Mott and Massey \cite{Mott-Massey}.
Note that our results are limited to the bare ion case. Therefore, an accurate comparison with experimental data requires consideration of all other screened ion projectiles, as reported in Ref. \cite{Cabrera-Trujillo-R-2023-98ps125401}, where the beam charge fraction is required.

For the coupled trajectories, the Bethe expression (Eq. (\ref{eq:bethe})) matches our $S_e$ results beyond the maxima. At very high collision energies, both the Bethe expression and the experimental data are in good agreement.
However, the $S_e$ results deviate from the Bethe expression for collision energies below the maxima, where the charge exchange process becomes significant (see Fig. \ref{fig:4}). For all projectiles, the electronic stopping cross section reaches a minimum near $E_p\approx0.1$ keV/u and generally peaks around $E_p\approx50$ keV/u. In the projectile energy range of 0.1 to $\sim$1 keV/u, nuclear target recoil becomes the dominant channel contributing to the total stopping cross section.

For the straight-line results (solid purple line), the best agreement with the coupled trajectories, Bethe, and experimental data occurs only at very high collision energies. In the remainder of the energy range, the electronic stopping cross sections derived from the straight-line approximations differ significantly from those obtained from the coupled trajectories.
For the H$^+$ projectile, the straight-line results reach a maximum (beyond the scale of our figure) at a much lower collision energy than the coupled results. This would imply that, at low energies, the target gains more energy than is physically available from the projectile, which violates energy conservation. For He$^{2+}$, Li$^{3+}$, and Be$^{4+}$ projectiles, the straight-line electronic stopping cross sections are smaller due to the more pronounced charge transfer process, which reduces the probability of finding the electron in the target (see Fig. \ref{fig:2}). This is a consequence of the higher projectile charge. Additionally, there are still regions where $S_e$ exceeds $S_T$, which is unphysical, particularly at low collision energies. Therefore, it is clear that there is a strong dependence on trajectory for energy and momentum transfer processes in these collisions. 

In Table \ref{tab:1}, we provide the numerical results from the electron-nuclei coupled trajectories for the electron capture, total, and nuclear cross sections for reference purposes.

\section{Conclusions}
\label{sec:conclusions}
In this work, our goal was to determine whether trajectory effects influence inelastic processes such as charge exchange and energy loss in collisions of bare ion projectiles with atomic hydrogen.
We used a lattice representation to numerically solve the time-dependent Schrödinger equation with a coupled electron-nuclear dynamics approach as well as a straight-line trajectory approximation for the study. This method provided excellent results for electron capture cross sections, which agree well with experimental and theoretical data. From this, we determined that the charge exchange process is unaffected by the trajectory. However, for energy loss, the straight-line trajectories lead to the projectile inducing a higher energy gain for the target, which is inconsistent with physical principles. Thus, this work asserts that a proper description of the electron-nuclear dynamics is necessary to account for the energy and momentum transfer processes in collisions of bare ions with atomic hydrogen when compared with simpler straight-line trajectory calculations.

We hope our results will encourage further interest from the experimental community to investigate stopping cross sections at low collision energies. Additionally, we intend for this work to be used as a benchmark for other theoretical methods, including classical, semi-classical, and perturbative approaches, particularly in the low energy range where agreement has been difficult to establish. Building on these findings, we aim to apply this method to study collisions involving more complex, multi-electronic systems. Specifically, this method can be applied to investigate charge exchange cross sections of negatively charged ions originating from a positive charge state. This is particularly relevant for low-yield radioactive species where producing ions in the negative charge state through electron capture reactions is the most viable approach.

\section*{Acknowledgments}
R.C.T. and M.N. acknowledge support from grants UNAM-DGAPA-PAPIIT IN-109-623 and LANCAD-UNAM-DGTIC-228, as well as to the University of Gothenburg for providing the conditions for a visiting stay. D.H. and M.N. acknowledge support from the Swedish Research Council (2020-03505). This project has received funding from the European Union’s Horizon 2020 research and innovation programme under the Marie Skłodowska-Curie grant agreement No 861198.

\bibliographystyle{apsrev4-1}
\input{text.bbl}

\end{document}

%% file: text.bbl
%